**Title**: Hofstadter Butterfly in Graphene

**Authors**: Wei Yang, Guangyu Zhang

**Affiliations**: Institute of Physics, Chinese Academy of Sciences, Beijing 100190, China

**Abstract**: This chapter reports recent progresses on the observation of Hofstadter butterfly in graphene, and organized as follows. We first briefly introduce the concept of Hofstadter butterfly, including the theoretical model of Bloch electrons in the magnetic field, the resulting Harper's equation and its solution, and its realizations in non-graphene systems. Then, we discuss the ways to realize the fractal Hofstadter spectra in graphene, and introduce three types of graphene superlattice structure, including graphene/hBN, twisted graphene layers, nanofabricated graphene superlattice. In particular, details about the fractal Hofstadter spectrum in graphene will be focused and discussed in the graphene/hBN superlattice and the twisted bilayer graphene.

**Key words**: Hofstadter Butterfly, Landau level, Quantum Hall effect, Quantum anomalous Hall effect, Graphene, moiré superlattice, twisted graphene multilayers, band topology, Chern insulator, Symmetry breaking phases, magneto-transport, quantum capacitance, scanning probe techniques.

**Key points**:

➢ Introduction of Hofstadter Butterfly and the realization in non-graphene systems

➢ Graphene superlattices: graphene/hBN, twisted graphene layers, nanofabricated graphene superlattice

➢ The realization of fractal Hofstadter butterfly in Graphene/hBN superlattice

➢ Topological Chern insulators and Hofstadter butterfly in Twisted bilayer graphene

**Introduction**

Hofstadter butterfly refers to the energy spectrum of two-dimensional (2D) electrons being subjected to both the periodic electrostatic potential and a perpendicular magnetic field (B), which leads to a self-similar recursive Landau level spectrum resembling a butterfly. It is named after D. Hofstadter who first obtained the fractal energy spectrum mathematically for a square lattice in 1976 (Hofstadter, 1976). He assumed that the tight-binding Bloch energy has the form of $2E_0(\cos k_x \lambda + \cos k_y \lambda)$, where $E_0$ is an empirical parameter, $\lambda$ is the electrostatic potential period, and **k** is the wave vector. He then replaced the momentum by doing a Peierls substitution, i.e. $\hbar\mathbf{k} \to \hbar\mathbf{k} - e\mathbf{A}$, where e is the electron charge and **A** is the vector potential with $\mathbf{B} = \nabla \times \mathbf{A}$. By solving the time-independent Schrodinger equation with the ansatz of $\psi(x,y) = g_n e^{i\alpha m}$, where $x = n\lambda$ and $y = m\lambda$ (n and m the integers) with $\alpha$ depends on energy, one could obtain Harper's equation:

$$g_{n+1} + g_{n-1} + 2\cos(2\pi n\phi/\phi_0 - \alpha) g_n = \epsilon g_n,$$

with $\phi = B\lambda^2$ is the magnetic flux through the unit cell, $\phi_0 = h/e$ is the magnetic quantum flux with $h$ the Planck constant, and $\epsilon = 2E/E_0$ with E the energy. At rational fillings with $\phi/\phi_0 = p/q$, where p and q are co-prime integers, the solution of the Harper's equation reveals that the Bloch band would split into $q$ distinct energy subbands, resulting in a recursive fractal energy diagram as shown Fig.1, the so called Hofstadter butterfly.

In 1978, G. Wannier (Wannier, 1978) took the density of state into account and transformed the fractal energy spectrum into a linear trajectory of density-field diagram, so called Wannier diagram. In this representation, the

Hofstadter butterfly is described by the Diophantine relation:

$$n/n_0 = t(\phi/\phi_0) + s$$

Here n is the carrier density and $n_0$ is the carrier density of a unit cell. The first quantum number *t* corresponds to the topological properties of Landau level flat band and the second quantum number *s* is the Bloch band filling index. Later on, the fractal spectrum was also calculated for triangular lattices (Claro and Wannier, 1979) and honeycomb lattices (Rammal, 1985).

To experimentally observe the Hofstadter butterfly is rather challenging. The minigaps of the fractal spectrum is considerable when magnetic flux is comparable to the quantum flux, in other words, the magnetic length $l_B = \sqrt{\hbar/eB}$ is of the same order as λ, the period of electrostatic potential of the Bloch band. For a real crystal, the period is the lattice constant, which is of the order of sub-nanometer, and thus it usually requires an extremely large magnetic field over 1000T that is experimentally inaccessible in the lab. Back in the 1990s, electrostatic periods much larger than its lattice constant were achieved by fabricating lateral superlattices for 2D electron gas in GaAs/AlGaAs heterostructures (Schlösser, *et al.*, 1996; Albrecht, *et al.*, 1999; Albrecht, *et al.*, 2001; Albrecht, *et al.*, 2003; Geisler, *et al.*, 2004). Although signatures of Hofstadter spectrum were observed in these artificial superlattices, the device quality degradation due to nanofabrication as well as the difficulty in tuning carrier density hindered the revelation of the fractal structure.

The fractal Hofstadter spectrum was vividly realized in a microwave waveguide with a periodic arrangement of scatterers (Kuhl and Stöckmann, 1998). In this photonic system, the transmission matrix mimics the Hamiltonian of Bloch electrons in the magnetic field, and equivalently leads to the Harper equation and a fractal Hofstadter butterfly. Similar realizations of the Hofstadter Hamiltonian were also reported in optical lattices with ultracold atoms (Aidelsburger, *et al.*, 2013) and in superconducting qubits (Roushan, *et al.*, 2017).

**Graphene superlattices:**

Being atomically thin, graphene is an ideal 2D electron system. Soon after the discovery of quantum Hall effect in graphene (Novoselov, *et al.*, 2005; Zhang, *et al.*, 2005), calculations of the fractal Hofstadter spectrum were carried out in monolayer graphene (Hasegawa and Kohmoto, 2006) and bilayer graphene (Nemec and Cuniberti, 2007). Yet, the problem is obvious as the graphene lattice constant is small, *a* = 0.246 nm, thus requiring a superlattice structure.

One solution is to realize graphene superlattice structures with external periodic potentials, whose band structure was calculated by C. Park et al. (Park, *et al.*, 2008), revealing an emergence of secondary Dirac point at the superlattice Brillouin zone boundary. Similar calculations were also done by L. Brey and H. Fertig 2009, and M. Barbier et al. 2010 (Brey and Fertig, 2009; Barbier, *et al.*, 2010). The first and also the best experimental demonstration is the moiré superlattice structure on graphene/hexagonal boron nitride (hBN) heterostructure, formed due to their lattice mismatch as illustrated in Fig. 2a. The graphene/hBN superlattice was initially observed in a scanning tunneling microscope/spectrum (STM/STS) study by M. Yankowitz et al. (Yankowitz, *et al.*, 2012). Perfect twist angle control with the $\theta = 0°$ was realized via direct epitaxy of graphene on hBN with a superlattice period of ~15.6nm (Yang, *et al.*, 2013). Note that the twist angle between graphene and hBN could also be tuned by thermal annealing (Wang, *et al.*, 2016) and AFM tip manipulation (Koren, *et al.*, 2016; Wang, *et al.*, 2016; Ribeiro-Palau, *et al.*, 2018). The fractal Hofstadter spectrum was realized in transferred graphene/hBN superlattices (Dean, *et al.*, 2013; Hunt, *et al.*, 2013; Ponomarenko, *et al.*, 2013; Yu, *et al.*, 2014; Wang, *et al.*, 2015; Krishna Kumar, *et al.*, 2017; Krishna

Kumar, *et al.*, 2018; Spanton, *et al.*, 2018), in epitaxial graphene/hBN superlattices (Yang, *et al.*, 2013; Yang, *et al.*, 2016; Chen, *et al.*, 2017; Lu, *et al.*, 2020), and in ABC-trilayer graphene/hBN superlattice (Chen, *et al.*, 2020a). Soon after the experimental observations, the fractal Hofstadter spectrum were calculated in graphene/hBN superlattice (Chen, *et al.*, 2014).

Besides, there are also other graphene superlattices based on nanofabrication, such as one-dimensional (1D) superlattices with a comb-like gate (Dubey, *et al.*, 2013; Drienovsky, *et al.*, 2014), etched graphene antidots (Sandner, *et al.*, 2015; Yagi, *et al.*, 2015), and patterned dielectric superlattice (Forsythe, *et al.*, 2018), as sketched in Fig. 2b. However, the results are not promising. The 1D gated superlattices suffered from serious p-n junction effects, and the graphene antidots were limited by its big superlattice size and thus dominated by classic commensurate effect between the superlattice period and the cyclotron orbits. The patterned dielectric superlattices with a reduced period of ~40 nm and a stronger tunability of the periodic potential revealed signatures of the fractal spectrum; however, the quality was still limited as it suffered from similar problems as those of lateral superlattices of the 2D electron gas in GaAs/AlGaAs heterostructures.

Another strategy in this effort is a moiré butterfly by rotating two graphene layers with the period of $\lambda = a/(2\sin(\theta/2))$ at a twisted angle ($\theta$) (Fig. 2c). This was firstly proposed in (Bistritzer and MacDonald, 2011b), where fractal spectrum at a strong field limit ($\phi/\phi_0 > 1$) was investigated. Later, calculations of the fractal spectrum at a weak field limit ($1 > \phi/\phi_0 > 0$) were carried out in twisted bilayer graphene (TBG) for different twist angles by P. Moon and M. Koshino and Z. Wang et al. (Moon and Koshino, 2012; Wang, *et al.*, 2012). In fact, the moiré superlattice structure of the TBG were observed in a series of the STM experiments (Li, *et al.*, 2010; Luican, *et al.*, 2011; Brihuega, *et al.*, 2012). In the early works of quantum transport on TBG prepared either by CVD growth on SiC (Lee, *et al.*, 2011) or by folded graphene (Schmidt, *et al.*, 2014), the device quality was much limited and no signature of the fractal band was observed. Thanks to the development of "pick-up" transfer technique (Wang, *et al.*, 2013) and subsequent 'tear and stack' technique using the same graphene flake (Kim, *et al.*, 2016a), accurate control of twist angle in TBG and high-quality devices were possible. In 2016, realizations of TBG with $\theta \sim 2^o$ (corresponding to $\lambda \sim 7nm$) were reported by Y. Kim et al. (Kim, *et al.*, 2016b) and Y. Cao et al. (Cao, *et al.*, 2016), where two sets of LLs were observed; one stemming from the main charge neutral point (CNP) and the other from the superlattice gaps (or called full filling n=4$n_0$). However, the two LLs intersect each other without revealing the recurring fractal structures. Then, the magic angle TBG with $\theta \sim 1.1^o$ (corresponding to $\lambda \sim 12.8nm$) was firstly realized by Y. Cao (Cao, *et al.*, 2018a; Cao, *et al.*, 2018b), where the LLs are observed to fan out of the main CNP and superlattice gaps, and later by (Lu, *et al.*, 2019; Yankowitz, *et al.*, 2019) with the LLs developed from half filling and quarter fillings. The fractal Hofstadter spectrum was also observed in other twisted graphene multilayer moiré systems including but not limited to the twisted double bilayer graphene (TDBG) (Burg, *et al.*, 2019; Cao, *et al.*, 2020; Liu, *et al.*, 2020; Shen, *et al.*, 2020), twisted monolayer bilayer graphene (TMBG) (Chen, *et al.*, 2020b; Polshyn, *et al.*, 2020; He, *et al.*, 2021; Polshyn, *et al.*, 2021; Xu, *et al.*, 2021), and twisted trilayer graphene (TTG) (Hao, *et al.*, 2021; Park, *et al.*, 2021b; Siriviboon, *et al.*, 2021).

**Hofstadter butterfly in Graphene/hBN superlattice**:

In graphene/hBN heterostructure, graphene and hBN share a similar triangular lattice with a mismatch of $\delta = \sim 1.7\%$, which gives a triangular moiré superlattice with a period of $\lambda = (1 + \delta)a/\sqrt{2(1 + \delta)(1 - cos\theta) + \delta^2}$ at a given twist angle ($\theta$). The area of the superlattice unit cell $A = \sqrt{3}\lambda^2/2$ defines the number of electron states per area of a fully filled Bloch band $n_0 = 1/A$. Here, the carrier density n can be largely tuned by gating n=$C_G V_G/e$, where $C_G$ ($V_G$) is the geometrical capacitance (gate voltage), and the magnetic flux is defined by $\phi = BA$. Then, the measured Landau fan diagram, a mapping of longitudinal resistance ($R_{xx}$) or Hall resistance ($R_{xy}$) to the change of $V_G$ and $B$, could be transformed into the Wannier diagram of the fractal Hofstadter butterfly spectrum, as shown in Fig. 3, following the Diophantine relation: $n/n_0 = t(\phi/\phi_0) + s$, where both t and s are integers in the single-particle picture.

At zero magnetic field, i.e. $\phi/\phi_0 = 0$, the $R_{xx}$ peak at s=0 corresponds to the charge neutral point (CNP) of graphene, while two satellite peaks developed at s=±4 are referred to the gap openings at the moiré mini-Brillouin zone boundary (or called secondary Dirac points in some literatures), and the number 4 also means the 4-fold degeneracy (spin and valley) of the Bloch band, i.e. 4 electrons/holes to fully filled one moiré unit cell. There is electron-hole asymmetry, and usually the superlattice effect is stronger for holes than that for electrons.

At s=0, the Diophantine relation is simplified to $n/n_0 = t(\phi/\phi_0)$. This is exactly the case of 2D free electrons/holes subjected to a perpendicular magnetic field, and the quantum number *t* has the same physical meaning as the filling factors of Landau levels in graphene where the Hall conductance is quantized to $te^2/h$. Let us take the monolayer (bilayer) graphene/hBN superlattice as an example. The states developed at t=4N+2 (4N for bilayer graphene) means that the Landau quantization gaps with 4-fold symmetry, while those at other integers are corresponding to the gaps with broken symmetry, with N=0, 1, 2…Landau level index (N=1, 2, 3…for bilayer graphene).

At s=±4, the Diophantine relation can be modified as $(n/n_0 \mp 4) = t(\phi/\phi_0)$. At a small magnetic field before the Landau quantization, the sign of $R_{xy}$ changes at s=±4, which is exactly the same as sign change of $R_{xy}$ at CNP. As the magnetic field increases further, features of the Landau level states are found fanning out from s=±4 with a slop of $t = (\phi/\phi_0)/(n/n_0 \mp 4)$, similar to those fanning out from CNP with a slope of t=$(\phi/\phi_0)/(n/n_0)$. The whole spectra stemming at s=±4 is like a replica of those at CNP, a signature of fractal Hofstadter butterfly. Note that the LLs stemmed from s=-4 show up at almost the same magnetic field as those at the CNP.

The most intriguing evidence of the fractal Hofstadter spectrum is revealed at the fractional filling $\phi/\phi_0 = p/q$. If *p* =1 as in most cases, the LLs from s=-4 and those from s=0 intersect at $\phi/\phi_0 = 1/q$. For instance, the LLs with t=2 from s=-4 will meet those with t=-2, -6, and -10 from s=0 at *q*=1, 2, and 3, respectively, as shown in Fig.3b (Hunt, *et al.*, 2013). At these crossing points, mini Hofstadter gaps open at $\phi/\phi_0 = 1/q$, and moreover LLs with different *t* are developed at an effective magnetic field $B_{eff} = \pm|B - B_{1/q}|$, whose traces at B=0 could be any integer *s* between 0 and -4. In principle, similar recursive fractal gaps should be found at other crossing points in the Wannier diagram, and the whole spectrum constructs the fractal Hofstadter butterfly. Note that the fractional *s* with an integer *t* (also named symmetry broken Chern insulator) as well as a fractional *t* with s=0 (fractional quantum Hall insulator) are observed in graphene/hBN superlattice (Wang, *et al.*, 2015). Moreover, gaps with both fractional *s* and fractional *t* (also named fractional Chern insulator) are revealed in bilayer graphene/hBN superlattices (Spanton, *et al.*, 2018), indicating fractional Hofstadter butterfly spectra in the many-body picture.

The fractal Hofstadter spectrum can be probed in conductance measurement (Dean, *et al.*, 2013; Hunt, *et al.*, 2013; Ponomarenko, *et al.*, 2013; Wang, *et al.*, 2015; Yang, *et al.*, 2016; Chen, *et al.*, 2017; Krishna Kumar, *et al.*, 2017) and also in capacitance measurements (Hunt, *et al.*, 2013; Yu, *et al.*, 2014; Spanton, *et al.*, 2018). Note that the fractal Hofstadter butterfly spectra can survive at elevated temperature up to 100 K (Krishna Kumar, *et al.*, 2017; Krishna Kumar, *et al.*, 2018), even though the fractal Hofstadter butterfly spectra are thermally smeared into Brown-Zak oscillations, i.e., the conductance oscillations in $\phi_0/\phi$.

**Hofstadter butterfly in Twisted bilayer graphene**

The situation in TBG are more complicated and interesting, because now the twisted angle ($\theta$) determines not only the size of the moiré superlattice $\lambda = a/(2\sin(\theta/2))$, but also the interlayer coupling. At a magic angle of ~1.05°, the resulting moiré bands are extremely flat with a bandwidth of ~10 meV, implying a reduced Fermi velocity (Bistritzer and MacDonald, 2011a). The moiré flat band in the magic angle TBG gives rise to the strong electron correlation effects, including the correlated insulators at half fillings and superconductivity nearby (Cao, *et al.*, 2018a; Cao, *et al.*, 2018b), as well as correlator insulators at quarter fillings (Lu, *et al.*, 2019; Yankowitz, *et al.*, 2019). In TBG, it is convenient to define the moiré band filling factor ($v = n/n_0$), corresponding to the number of carriers filled in the moiré conduction/valence band, with $n_0 = 1/A$ and $A = \sqrt{3}\lambda^2/2$. As a result, the resistance peaks at $v=\pm 1$ ($\pm 3$), $\pm 2$, $\pm 4$ in the transfer curve are corresponding to the correlated insulators at quarter (three quarters) fillings, half fillings, and moiré band insulators, respectively.

The fractal Hofstadter butterfly spectrum in TBG is also revealed in the Wannier diagram, following the Diophantine relation: $n/n_0 = t(\phi/\phi_0) + s$, where both t and s are integers in the single-particle picture. The strong electron-electron interactions in TBG could substantially reshape the fractal Hofstadter spectrum. Usually, the correlated insulators are quite robust against the magnetic field, and the replica of the LLs emerges at these integer fillings $v$, unlike the dominating LLs at $v=0$ and $\pm 4$ in graphene/hBN superlattice. Note that the LL filling factor $v_{LL}$ in the replica spectrum at non-zero integer $v$ is usually two-fold or even one-fold degenerate, and the filling factors $v_{LL}$ have the same sign as those of $v$. In addition, the quantum oscillations in TBG are usually much weaker than those in the graphene/hBN superlattice. This is due to the reduced Fermi velocity of the flat band, especially for MATBG, and therefore the Landau quantization gaps as well as the fractal Hofstadter gaps are smaller compared to the dispersive moiré bands in the graphene/hBN superlattice.

Additionally, these moiré flat bands in TBG are topological with non-trivial Berry curvatures (Liu, *et al.*, 2019a; Liu, *et al.*, 2019b) at zero magnetic fields. The zero-field topological bands are mathematically the same as those LLs or fractal LLs at finite magnetic fields. Note that the quantum Hall insulator is in principle a Chern insulator where the LL filling factor corresponds to the Chern number, and here we use "Chern" specifically for the moiré Chern bands to avoid confusion. Considering spin, valley, and the moiré valley, the moiré bands are 8-fold degenerate at the Dirac point, protected by the $C_2T$ symmetry where $C_2$ is the inversion symmetry and $T$ is time reversal symmetry.

If $C_2T$ symmetry is unbroken, then the total Chern number is zero since the moiré bands with opposite Chern numbers touches at the Dirac point, even though each of the moiré bands are topological with C=±1. This can be related to the fragile topology (Ahn, *et al.*, 2019; Po, *et al.*, 2019; Song, *et al.*, 2019; Lian, *et al.*, 2020), whose signature is revealed in the observation of the C=0 band gaps being intersected by the non-zero LLs at $\phi/\phi_0 = 2$ (B ~9T) in the TBG with θ ~0.45º (Lu, *et al.*, 2021).

If the inversion symmetry is broken by a staggered potential mass while the *T* symmetry is preserved, it leads to C=+1 at the K valley and C=-1 at the K' valley in the moiré conduction band, and C=-1 at the K valley and C=+1 at the K' valley in the moiré valence band. As a result, the Chern insulator with C=+1 or -1 emerge when the odd number of carriers is filled in the moiré band. The breaking of the $C_2$ symmetry happens when the TBG is aligned with the hBN lattice, first realized by A. Sharpe (Sharpe, *et al.*, 2019) with the observations of ferromagnetism and Chern insulator with C=1 at *v*=3, and then by M. Serlin (Serlin, *et al.*, 2020) with the observation of quantized anomalous Hall effect (QAH) with C=1 at v=3, as shown Fig. 4a. Similar observations have also been observed in ABC-trilayer graphene/hBN superlattice (Chen, *et al.*, 2020a), twisted monolayer bilayer graphene (Chen, *et al.*, 2020b; Polshyn, *et al.*, 2020). These observations are in stark contrast to the conventional fractal Hofstadter butterfly that strongly depends on the magnetic flux in the moiré superlattice.

If the *T* symmetry is broken by the electron-electron interaction induced Haldane mass (Haldane, 1988), it leads to C=+1 at both the K and K' valley in the moiré conduction band, and C=-1 at the K and K' valley in the moiré valence band. As a result, it gives a Chern insulator with Chern number depending on a sequential filling of carriers, i.e. C = -4 - *v* for holes and C = 4 - *v* for electrons (as shown Fig. 4b). The Chern insulators with the *T* symmetry-broken are observed in various experiments, for instance in transport measurements (Das, *et al.*, 2021; Park, *et al.*, 2021a; Saito, *et al.*, 2021; Shen, *et al.*, 2021; Wu, *et al.*, 2021b) and scanning spectroscopic measurements (Nuckolls, *et al.*, 2020; Choi, *et al.*, 2021; Pierce, *et al.*, 2021). These Chern insulators are found to compete with the zero-field ground states in MATBG (Stepanov, *et al.*, 2021). These Chern insulators are also observed for the non-magic angle TBG fractal Hofstadter spectra at finite magnetic fields (Choi, *et al.*, 2021; Shen, *et al.*, 2021).

Moreover, fractional Chern insulators with fractional C and *v* (referred to the fractional *t* and fractional *s* in the Diophantine language), symmetry-broken Chern insulators (non-zero integers *t* and fractional *s*), and charge density waves (*t=0* and fractional *s*) are observed in local compressibility measurements (Xie, *et al.*, 2021), and are also studied in the numerical calculations (Andrews and Soluyanov, 2020; Wang and Santos, 2020). Similar topological phases with a fractional *v* are also observed in the twisted monolayer-bilayer graphene (Polshyn, *et al.*, 2021) and twisted trilayer graphene (Siriviboon, *et al.*, 2021).

All these Chern insulators as well as CDW phases are weakly dependent on the filling of the magnetic flux, unlike the fractal and fractional fractal Hofstadter spectra in graphene/hBN superlattices. In fact, it is also possible to suppress the Chern insulators and recover the conventional fractal Hofstadter spectrum by twisting away from the magic angle (Shen, *et al.*, 2021). Because the moiré bands are more dispersive for the non-magic angle, which leads to stronger Landau quantization (bigger fractal Hofstadter gaps) and weaker electron interactions (smaller Chern gaps). Also note that the moiré bands in TBG at a non-magical angle might be topologically trivial. Besides, it is also possible to reveal the fractal Hofstadter spectra in the MATBG by tuning the Fermi level from the flat moiré bands to the dispersive remote bands (Das, *et al.*, 2021). Last but not least, aside from TBG, other twisted graphene multilayer system offers a good opportunity to study the electrical field tunable Hofstadter spectrum. Take the TDBG for example, the Landau fan diagram of the Hofstadter butterfly spectrum is strongly displacement field tunable(Liu, *et al.*, 2022a) and stacking order dependent (Crosse, *et al.*, 2020; Wu, *et al.*, 2021a), and the valley polarized correlated insulators at half fillings in TDBG are observed topological(Liu, *et al.*, 2022b) and could host insulating quantum oscillations (Liu, *et al.*, 2022a). These observations suggest a rich interplay among the moiré flat band topology, symmetry, Landau quantization and others.

**Conclusions:**

Graphene is an ideal platform to experimentally realize the fractal Hofstadter butterfly. The development of van der Waals stacking techniques allows the realizations of high-quality graphene superlattice structures, including the graphene/hBN superlattice and the TBG and other twisted graphene multilayers. In a graphene/hBN superlattice, the recursive nature of the fractal Hofstadter spectra has been vividly revealed at rational fillings of $\phi/\phi_0$. In TBG, the fractal Hofstadter spectra is strongly twist-angle dependent. In particular, the magic angle TBG reveals abundant topological Chern insulators including the quantized anomalous Hall effect at zero magnetic field, which is distinct from the conventional fractal Hofstadter spectrum.

**References:**
Ahn, J., S. Park, and B.-J. Yang, 2019, "Failure of Nielsen-Ninomiya Theorem and Fragile Topology in Two-Dimensional Systems with Space-Time Inversion Symmetry: Application to Twisted Bilayer Graphene at Magic Angle," Physical Review X **9**, 021013.
Aidelsburger, M., M. Atala, M. Lohse, J. T. Barreiro, B. Paredes, and I. Bloch, 2013, "Realization of the Hofstadter Hamiltonian with ultracold atoms in optical lattices," Phys. Rev. Lett. **111**, 185301.
Albrecht, C., J. Smet, D. Weiss, K. Von Klitzing, R. Hennig, M. Langenbuch, M. Suhrke, U. Rössler, V. Umansky, and H. Schweizer, 1999, "Fermiology of two-dimensional lateral superlattices," Physical review letters **83**, 2234.
Albrecht, C., J. H. Smet, K. von Klitzing, D. Weiss, V. Umansky, and H. Schweizer, 2003, "Evidence of Hofstadter's fractal energy spectrum in the quantized Hall conductance," Physica E: Low-dimensional Systems and Nanostructures **20**, 143-148.
Albrecht, C., J. H. Smet, K. von Klitzing, D. Weiss, V. V. Umansky, and H. Schweizer, 2001, "Evidence of Hofstadter's Fractal Energy Spectrum in the Quantized Hall Conductance," Phys Rev Lett **86**, 147-150.
Andrews, B., and A. Soluyanov, 2020, "Fractional quantum Hall states for moir\'e superstructures in the Hofstadter regime," Physical Review B **101**, 235312.
Barbier, M., P. Vasilopoulos, and F. M. Peeters, 2010, "Extra Dirac points in the energy spectrum for superlattices on single-layer graphene," Physical Review B **81**, 075438.
Bistritzer, R., and A. H. MacDonald, 2011a, "Moire bands in twisted double-layer graphene," Proceedings of the National Academy of Sciences of the United States of America **108**, 12233-12237.
Bistritzer, R., and A. H. MacDonald, 2011b, "Moiré butterflies in twisted bilayer graphene," Physical Review B **84**, 035440.
Brey, L., and H. A. Fertig, 2009, "Emerging zero modes for graphene in a periodic potential," Phys Rev Lett **103**, 046809.
Brihuega, I., P. Mallet, H. González-Herrero, G. Trambly de Laissardière, M. M. Ugeda, L. Magaud, J. M. Gómez-Rodríguez, F. Ynduráin, and J. Y. Veuillen, 2012, "Unraveling the Intrinsic and Robust Nature of van Hove Singularities in Twisted Bilayer Graphene by Scanning Tunneling Microscopy and Theoretical Analysis," Physical Review Letters **109**, 196802.
Burg, G. W., J. Zhu, T. Taniguchi, K. Watanabe, A. H. MacDonald, and E. Tutuc, 2019, "Correlated Insulating States in Twisted Double Bilayer Graphene," Physical Review Letters **123**, 197702.
Cao, Y., V. Fatemi, A. Demir, S. Fang, S. L. Tomarken, J. Y. Luo, J. D. Sanchez-Yamagishi, K. Watanabe, T. Taniguchi, E. Kaxiras, *et al.*, 2018a, "Correlated insulator behaviour at half-filling in magic-angle graphene superlattices," Nature **556**, 80-84.

Cao, Y., V. Fatemi, S. Fang, K. Watanabe, T. Taniguchi, E. Kaxiras, and P. Jarillo-Herrero, 2018b, "Unconventional superconductivity in magic-angle graphene superlattices," Nature **556**, 43-50.

Cao, Y., J. Y. Luo, V. Fatemi, S. Fang, J. D. Sanchez-Yamagishi, K. Watanabe, T. Taniguchi, E. Kaxiras, and P. Jarillo-Herrero, 2016, "Superlattice-Induced Insulating States and Valley-Protected Orbits in Twisted Bilayer Graphene," Physical Review Letters **117**, 116804.

Cao, Y., D. Rodan-Legrain, O. Rubies-Bigorda, J. M. Park, K. Watanabe, T. Taniguchi, and P. Jarillo-Herrero, 2020, "Tunable correlated states and spin-polarized phases in twisted bilayer–bilayer graphene," Nature **583**, 215-220.

Chen, G., A. L. Sharpe, E. J. Fox, Y.-H. Zhang, S. Wang, L. Jiang, B. Lyu, H. Li, K. Watanabe, T. Taniguchi, *et al.*, 2020a, "Tunable correlated Chern insulator and ferromagnetism in a moire superlattice," Nature **579**, 56-61.

Chen, G., M. Sui, D. Wang, S. Wang, J. Jung, P. Moon, S. Adam, K. Watanabe, T. Taniguchi, S. Zhou, *et al.*, 2017, "Emergence of Tertiary Dirac Points in Graphene Moire Superlattices," Nano Lett **17**, 3576-3581.

Chen, S., M. He, Y.-H. Zhang, V. Hsieh, Z. Fei, K. Watanabe, T. Taniguchi, D. H. Cobden, X. Xu, C. R. Dean, *et al.*, 2020b, "Electrically tunable correlated and topological states in twisted monolayer–bilayer graphene," Nature Physics **17**, 374-380.

Chen, X., J. R. Wallbank, A. A. Patel, M. Mucha-Kruczyński, E. McCann, and V. I. Fal'ko, 2014, "Dirac edges of fractal magnetic minibands in graphene with hexagonal moiré superlattices," Physical Review B **89**, 075401.

Choi, Y., H. Kim, Y. Peng, A. Thomson, C. Lewandowski, R. Polski, Y. Zhang, H. S. Arora, K. Watanabe, T. Taniguchi, *et al.*, 2021, "Correlation-driven topological phases in magic-angle twisted bilayer graphene," Nature **589**, 536-541.

Claro, F. H., and G. H. Wannier, 1979, "Magnetic subband structure of electrons in hexagonal lattices," Physical Review B **19**, 6068-6074.

Crosse, J. A., N. Nakatsuji, M. Koshino, and P. Moon, 2020, "Hofstadter butterfly and the quantum Hall effect in twisted double bilayer graphene," Physical Review B **102**, 035421.

Das, I., X. Lu, J. Herzog-Arbeitman, Z.-D. Song, K. Watanabe, T. Taniguchi, B. A. Bernevig, and D. K. Efetov, 2021, "Symmetry-broken Chern insulators and Rashba-like Landau-level crossings in magic-angle bilayer graphene," Nature Physics **17**, 710-714.

Dean, C. R., L. Wang, P. Maher, C. Forsythe, F. Ghahari, Y. Gao, J. Katoch, M. Ishigami, P. Moon, M. Koshino, *et al.*, 2013, "Hofstadter's butterfly and the fractal quantum Hall effect in moire superlattices," Nature **497**, 598-602.

Drienovsky, M., F.-X. Schrettenbrunner, A. Sandner, D. Weiss, J. Eroms, M.-H. Liu, F. Tkatschenko, and K. Richter, 2014, "Towards superlattices: Lateral bipolar multibarriers in graphene," Physical Review B **89**, 115421.

Dubey, S., V. Singh, A. K. Bhat, P. Parikh, S. Grover, R. Sensarma, V. Tripathi, K. Sengupta, and M. M. Deshmukh, 2013, "Tunable superlattice in graphene to control the number of Dirac points," Nano Lett **13**, 3990-3995.

Forsythe, C., X. Zhou, K. Watanabe, T. Taniguchi, A. Pasupathy, P. Moon, M. Koshino, P. Kim, and C. R. Dean, 2018, "Band structure engineering of 2D materials using patterned dielectric superlattices," Nat Nanotechnol **13**, 566-571.

Geisler, M. C., J. H. Smet, V. Umansky, K. von Klitzing, B. Naundorf, R. Ketzmerick, and H. Schweizer, 2004, "Detection of a Landau band-coupling-induced rearrangement of the Hofstadter butterfly," Phys Rev Lett **92**, 256801.

Haldane, F. D. M., 1988, "Model for a Quantum Hall Effect without Landau Levels: Condensed-Matter Realization of the "Parity Anomaly"," Physical Review Letters **61**, 2015-2018.

Hao, Z., A. Zimmerman, P. Ledwith, E. Khalaf, D. H. Najafabadi, K. Watanabe, T. Taniguchi, A. Vishwanath, and P. Kim, 2021, "Electric field–tunable superconductivity in alternating-twist magic-angle trilayer graphene," Science **371**, 1133-1138.

Hasegawa, Y., and M. Kohmoto, 2006, "Quantum Hall effect and the topological number in graphene," Physical Review B **74**, 155415.

He, M., Y. H. Zhang, Y. Li, Z. Fei, K. Watanabe, T. Taniguchi, X. Xu, and M. Yankowitz, 2021, "Competing correlated states and abundant orbital magnetism in twisted monolayer-bilayer graphene," Nat Commun **12**, 4727.

Hofstadter, D. R., 1976, "Energy levels and wave functions of Bloch electrons in rational and irrational magnetic fields,"

Physical Review B **14**, 2239-2249.

Hunt, B., J. D. Sanchez-Yamagishi, A. F. Young, M. Yankowitz, B. J. LeRoy, K. Watanabe, T. Taniguchi, P. Moon, M. Koshino, P. Jarillo-Herrero*, et al.*, 2013, "Massive Dirac Fermions and Hofstadter Butterfly in a van der Waals Heterostructure," Science **340**, 1427-1430.

Kim, K., M. Yankowitz, B. Fallahazad, S. Kang, H. C. P. Movva, S. Huang, S. Larentis, C. M. Corbet, T. Taniguchi, K. Watanabe*, et al.*, 2016a, "van der Waals Heterostructures with High Accuracy Rotational Alignment," Nano Letters **16**, 1989-1995.

Kim, Y., P. Herlinger, P. Moon, M. Koshino, T. Taniguchi, K. Watanabe, and J. H. Smet, 2016b, "Charge Inversion and Topological Phase Transition at a Twist Angle Induced van Hove Singularity of Bilayer Graphene," Nano Lett **16**, 5053-5059.

Koren, E., I. Leven, E. Lortscher, A. Knoll, O. Hod, and U. Duerig, 2016, "Coherent commensurate electronic states at the interface between misoriented graphene layers," Nat Nanotechnol **11**, 752-7.

Krishna Kumar, R., X. Chen, G. Auton, A. Mishchenko, D. A. Bandurin, S. V. Morozov, Y. Cao, E. Khestanova, M. Ben Shalom, A. Kretinin*, et al.*, 2017, "High-temperature quantum oscillations caused by recurring Bloch states in graphene superlattices," Science **357**, 181-184.

Krishna Kumar, R., A. Mishchenko, X. Chen, S. Pezzini, G. H. Auton, L. A. Ponomarenko, U. Zeitler, L. Eaves, V. I. Fal'ko, and A. K. Geim, 2018, "High-order fractal states in graphene superlattices," Proc Natl Acad Sci U S A **115**, 5135-5139.

Kuhl, U., and H.-J. Stöckmann, 1998, "Microwave realization of the Hofstadter butterfly," Physical review letters **80**, 3232.

Lee, D. S., C. Riedl, T. Beringer, A. H. Castro Neto, K. von Klitzing, U. Starke, and J. H. Smet, 2011, "Quantum Hall effect in twisted bilayer graphene," Phys Rev Lett **107**, 216602.

Li, G., A. Luican, J. M. B. Lopes dos Santos, A. H. Castro Neto, A. Reina, J. Kong, and E. Y. Andrei, 2010, "Observation of Van Hove singularities in twisted graphene layers," Nature Physics **6**, 109-113.

Lian, B., F. Xie, and B. A. Bernevig, 2020, "Landau level of fragile topology," Physical Review B **102**, 041402.

Liu, J., J. Liu, and X. Dai, 2019a, "Pseudo Landau level representation of twisted bilayer graphene: Band topology and implications on the correlated insulating phase," Physical Review B **99**, 155415.

Liu, J., Z. Ma, J. Gao, and X. Dai, 2019b, "Quantum Valley Hall Effect, Orbital Magnetism, and Anomalous Hall Effect in Twisted Multilayer Graphene Systems," Physical Review X **9**, 031021.

Liu, L., Y. Chu, G. Yang, Y. Yuan, F. Wu, Y. Ji, J. Tian, R. Yang, K. Watanabe, and T. Taniguchi, 2022a, "Quantum oscillations in correlated insulators of a moir\'e superlattice," arXiv preprint arXiv:2205.10025.

Liu, L., S. Zhang, Y. Chu, C. Shen, Y. Huang, Y. Yuan, J. Tian, J. Tang, Y. Ji, R. Yang*, et al.*, 2022b, "Isospin competitions and valley polarized correlated insulators in twisted double bilayer graphene," Nature Communications **13**, 3292.

Liu, X., Z. Hao, E. Khalaf, J. Y. Lee, Y. Ronen, H. Yoo, D. Haei Najafabadi, K. Watanabe, T. Taniguchi, A. Vishwanath*, et al.*, 2020, "Tunable spin-polarized correlated states in twisted double bilayer graphene," Nature **583**, 221-225.

Lu, X., B. Lian, G. Chaudhary, B. A. Piot, G. Romagnoli, K. Watanabe, T. Taniguchi, M. Poggio, A. H. MacDonald, B. A. Bernevig*, et al.*, 2021, "Multiple flat bands and topological Hofstadter butterfly in twisted bilayer graphene close to the second magic angle," Proc Natl Acad Sci U S A **118**, e2100006118.

Lu, X., P. Stepanov, W. Yang, M. Xie, M. A. Aamir, I. Das, C. Urgell, K. Watanabe, T. Taniguchi, G. Zhang*, et al.*, 2019, "Superconductors, orbital magnets and correlated states in magic-angle bilayer graphene," Nature **574**, 653-657.

Lu, X., J. Tang, J. R. Wallbank, S. Wang, C. Shen, S. Wu, P. Chen, W. Yang, J. Zhang, K. Watanabe*, et al.*, 2020, "High-order minibands and interband Landau level reconstruction in graphene moiré superlattices," Physical Review B **102**, 045409.

Luican, A., G. Li, A. Reina, J. Kong, R. R. Nair, K. S. Novoselov, A. K. Geim, and E. Y. Andrei, 2011, "Single-Layer Behavior and Its Breakdown in Twisted Graphene Layers," Physical Review Letters **106**, 126802.

Moon, P., and M. Koshino, 2012, "Energy spectrum and quantum Hall effect in twisted bilayer graphene," Physical Review B **85**, 195458.

Nemec, N., and G. Cuniberti, 2007, "Hofstadter butterflies of bilayer graphene," Physical Review B **75**, 201404.

Novoselov, K. S., A. K. Geim, S. V. Morozov, D. Jiang, M. I. Katsnelson, I. V. Grigorieva, S. V. Dubonos, and A. A. Firsov, 2005, "Two-dimensional gas of massless Dirac fermions in graphene," Nature **438**, 197-200.

Nuckolls, K. P., M. Oh, D. Wong, B. Lian, K. Watanabe, T. Taniguchi, B. A. Bernevig, and A. Yazdani, 2020, "Strongly correlated Chern insulators in magic-angle twisted bilayer graphene," Nature **588**, 610-615.

Park, C. H., L. Yang, Y. W. Son, M. L. Cohen, and S. G. Louie, 2008, "New generation of massless Dirac fermions in graphene under external periodic potentials," Phys Rev Lett **101**, 126804.

Park, J. M., Y. Cao, K. Watanabe, T. Taniguchi, and P. Jarillo-Herrero, 2021a, "Flavour Hund's coupling, Chern gaps and charge diffusivity in moire graphene," Nature **592**, 43-48.

Park, J. M., Y. Cao, K. Watanabe, T. Taniguchi, and P. Jarillo-Herrero, 2021b, "Tunable strongly coupled superconductivity in magic-angle twisted trilayer graphene," Nature **590**, 249-255.

Pierce, A. T., Y. Xie, J. M. Park, E. Khalaf, S. H. Lee, Y. Cao, D. E. Parker, P. R. Forrester, S. Chen, K. Watanabe*, et al.*, 2021, "Unconventional sequence of correlated Chern insulators in magic-angle twisted bilayer graphene," Nature Physics **17**, 1210-1215.

Po, H. C., L. Zou, T. Senthil, and A. Vishwanath, 2019, "Faithful tight-binding models and fragile topology of magic-angle bilayer graphene," Physical Review B **99**, 195455.

Polshyn, H., Y. Zhang, M. A. Kumar, T. Soejima, P. Ledwith, K. Watanabe, T. Taniguchi, A. Vishwanath, M. P. Zaletel, and A. F. Young, 2021, "Topological charge density waves at half-integer filling of a moiré superlattice," Nature Physics **18**, 42-47.

Polshyn, H., J. Zhu, M. A. Kumar, Y. Zhang, F. Yang, C. L. Tschirhart, M. Serlin, K. Watanabe, T. Taniguchi, A. H. MacDonald*, et al.*, 2020, "Electrical switching of magnetic order in an orbital Chern insulator," Nature **588**, 66-70.

Ponomarenko, L. A., R. V. Gorbachev, G. L. Yu, D. C. Elias, R. Jalil, A. A. Patel, A. Mishchenko, A. S. Mayorov, C. R. Woods, J. R. Wallbank*, et al.*, 2013, "Cloning of Dirac fermions in graphene superlattices," Nature **497**, 594-597.

Rammal, R., 1985, "Landau level spectrum of Bloch electrons in a honeycomb lattice," Journal de Physique **46**, 1345-1354.

Ribeiro-Palau, R., C. Zhang, K. Watanabe, T. Taniguchi, J. Hone, and C. R. Dean, 2018, "Twistable electronics with dynamically rotatable heterostructures," Science **361**, 690-693.

Roushan, P., C. Neill, J. Tangpanitanon, V. M. Bastidas, A. Megrant, R. Barends, Y. Chen, Z. Chen, B. Chiaro, A. Dunsworth*, et al.*, 2017, "Spectroscopic signatures of localization with interacting photons in superconducting qubits," Science **358**, 1175-1179.

Saito, Y., J. Ge, L. Rademaker, K. Watanabe, T. Taniguchi, D. A. Abanin, and A. F. Young, 2021, "Hofstadter subband ferromagnetism and symmetry-broken Chern insulators in twisted bilayer graphene," Nature Physics **17**, 478-481.

Sandner, A., T. Preis, C. Schell, P. Giudici, K. Watanabe, T. Taniguchi, D. Weiss, and J. Eroms, 2015, "Ballistic Transport in Graphene Antidot Lattices," Nano Lett **15**, 8402-8406.

Schlösser, T., K. Ensslin, J. P. Kotthaus, and M. Holland, 1996, "Landau subbands generated by a lateral electrostatic superlattice-chasing the Hofstadter butterfly," Semiconductor Science Technology **11**, 1582.

Schmidt, H., J. C. Rode, D. Smirnov, and R. J. Haug, 2014, "Superlattice structures in twisted bilayers of folded graphene," Nature Communications **5**, 5742.

Serlin, M., C. L. Tschirhart, H. Polshyn, Y. Zhang, J. Zhu, K. Watanabe, T. Taniguchi, L. Balents, and A. F. Young, 2020, "Intrinsic quantized anomalous Hall effect in a moire heterostructure," Science **367**, 900-903.

Sharpe, A. L., E. J. Fox, A. W. Barnard, J. Finney, K. Watanabe, T. Taniguchi, M. A. Kastner, and D. Goldhaber-Gordon, 2019, "Emergent ferromagnetism near three-quarters filling in twisted bilayer graphene," Science **365**, 605-608.

Shen, C., Y. Chu, Q. Wu, N. Li, S. Wang, Y. Zhao, J. Tang, J. Liu, J. Tian, K. Watanabe*, et al.*, 2020, "Correlated states in twisted double bilayer graphene," Nature Physics **16**, 520-525.

Shen, C., J. Ying, L. Liu, J. Liu, N. Li, S. Wang, J. Tang, Y. Zhao, Y. Chu, K. Watanabe*, et al.*, 2021, "Emergence of Chern Insulating States in Non-Magic Angle Twisted Bilayer Graphene," Chinese Physics Letters **38**, 047301.

Sirivboon, P., J.-X. Lin, H. D. Scammell, S. Liu, D. Rhodes, K. Watanabe, T. Taniguchi, J. Hone, M. S. Scheurer, and J. Li, 2021,

"Abundance of density wave phases in twisted trilayer graphene on WSe$_2$," arXiv preprint arXiv:2112.07127.

Song, Z., Z. Wang, W. Shi, G. Li, C. Fang, and B. A. Bernevig, 2019, "All Magic Angles in Twisted Bilayer Graphene are Topological," Physical Review Letters **123**, 036401.

Spanton, E. M., A. A. Zibrov, H. Zhou, T. Taniguchi, K. Watanabe, M. P. Zaletel, and A. F. Young, 2018, "Observation of fractional Chern insulators in a van der Waals heterostructure," Science **360**, 62-66.

Stepanov, P., M. Xie, T. Taniguchi, K. Watanabe, X. Lu, A. H. MacDonald, B. A. Bernevig, and D. K. Efetov, 2021, "Competing Zero-Field Chern Insulators in Superconducting Twisted Bilayer Graphene," Phys Rev Lett **127**, 197701.

Wang, D., G. Chen, C. Li, M. Cheng, W. Yang, S. Wu, G. Xie, J. Zhang, J. Zhao, X. Lu*, et al.*, 2016, "Thermally Induced Graphene Rotation on Hexagonal Boron Nitride," Phys Rev Lett **116**, 126101.

Wang, J., and L. H. Santos, 2020, "Classification of Topological Phase Transitions and van Hove Singularity Steering Mechanism in Graphene Superlattices," Physical Review Letters **125**, 236805.

Wang, L., Y. Gao, B. Wen, Z. Han, T. Taniguchi, K. Watanabe, M. Koshino, J. Hone, and C. R. Dean, 2015, "Evidence for a fractional fractal quantum Hall effect in graphene superlattices," Science **350**, 1231-1234.

Wang, L., I. Meric, P. Huang, Q. Gao, Y. Gao, H. Tran, T. Taniguchi, K. Watanabe, L. Campos, D. Muller*, et al.*, 2013, "One-dimensional electrical contact to a two-dimensional material," Science **342**, 614-617.

Wang, Z. F., F. Liu, and M. Y. Chou, 2012, "Fractal Landau-level spectra in twisted bilayer graphene," Nano Lett **12**, 3833-3838.

Wannier, G., 1978, "A result not dependent on rationality for Bloch electrons in a magnetic field," physica status solidi **88**, 757-765.

Wu, Q., J. Liu, Y. Guan, and O. V. Yazyev, 2021a, "Landau Levels as a Probe for Band Topology in Graphene Moir\'e Superlattices," Physical Review Letters **126**, 056401.

Wu, S., Z. Zhang, K. Watanabe, T. Taniguchi, and E. Y. Andrei, 2021b, "Chern insulators, van Hove singularities and topological flat bands in magic-angle twisted bilayer graphene," Nat Mater **20**, 488-494.

Xie, Y., A. T. Pierce, J. M. Park, D. E. Parker, E. Khalaf, P. Ledwith, Y. Cao, S. H. Lee, S. Chen, P. R. Forrester*, et al.*, 2021, "Fractional Chern insulators in magic-angle twisted bilayer graphene," Nature **600**, 439-443.

Xu, S., M. M. Al Ezzi, N. Balakrishnan, A. Garcia-Ruiz, B. Tsim, C. Mullan, J. Barrier, N. Xin, B. A. Piot, T. Taniguchi*, et al.*, 2021, "Tunable van Hove singularities and correlated states in twisted monolayer–bilayer graphene," Nature Physics **17**, 619-626.

Yagi, R., R. Sakakibara, R. Ebisuoka, J. Onishi, K. Watanabe, T. Taniguchi, and Y. Iye, 2015, "Ballistic transport in graphene antidot lattices," Physical Review B **92**, 195406.

Yang, W., G. Chen, Z. Shi, C.-C. Liu, L. Zhang, G. Xie, M. Cheng, D. Wang, R. Yang, D. Shi*, et al.*, 2013, "Epitaxial growth of single-domain graphene on hexagonal boron nitride," Nature Materials **12**, 792-797.

Yang, W., X. Lu, G. Chen, S. Wu, G. Xie, M. Cheng, D. Wang, R. Yang, D. Shi, and K. Watanabe, 2016, "Hofstadter butterfly and many-body effects in epitaxial graphene superlattice," Nano Letters **16**, 2387-2392.

Yankowitz, M., S. Chen, H. Polshyn, Y. Zhang, K. Watanabe, T. Taniguchi, D. Graf, A. F. Young, and C. R. Dean, 2019, "Tuning superconductivity in twisted bilayer graphene," Science **363**, 1059-1064.

Yankowitz, M., J. Xue, D. Cormode, J. D. Sanchez-Yamagishi, K. Watanabe, T. Taniguchi, P. Jarillo-Herrero, P. Jacquod, and B. J. LeRoy, 2012, "Emergence of superlattice Dirac points in graphene on hexagonal boron nitride," Nature Physics **8**, 382-386.

Yu, G. L., R. V. Gorbachev, J. S. Tu, A. V. Kretinin, Y. Cao, R. Jalil, F. Withers, L. A. Ponomarenko, B. A. Piot, M. Potemski*, et al.*, 2014, "Hierarchy of Hofstadter states and replica quantum Hall ferromagnetism in graphene superlattices," Nature Physics **10**, 525-529.

Zhang, Y., Y.-W. Tan, H. L. Stormer, and P. Kim, 2005, "Experimental observation of the quantum Hall effect and Berry's phase in graphene," Nature **438**, 201-204.

**Figures and Figure captions**

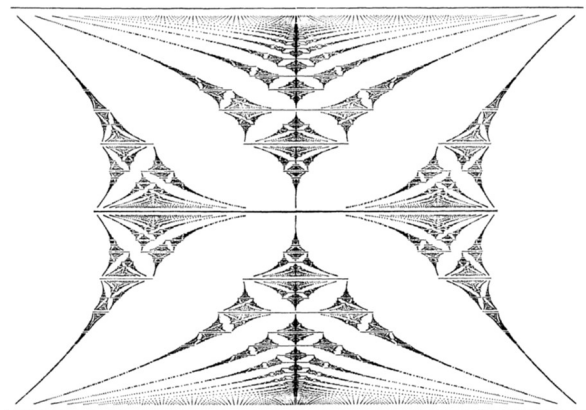

**Fig. 1 Calculated Fractal spectrum in a unit cell by D. Hofstadter, reproduced from (Hofstadter, 1976).**

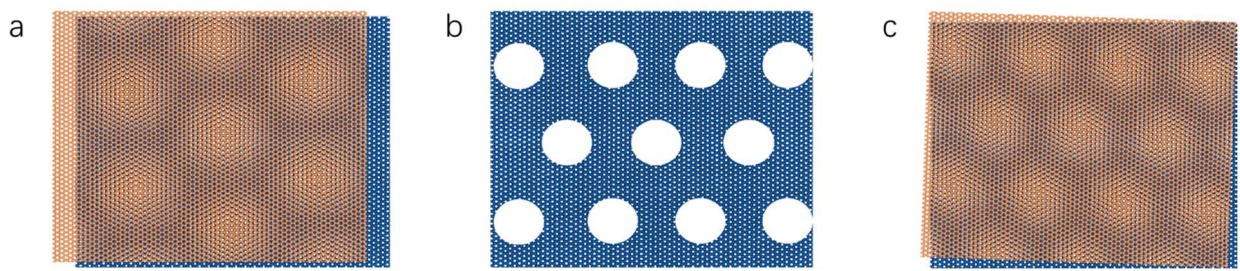

**Fig. 2 Sketches of graphene superlattice structures formed by lattice constant mismatch (a), nanofabrication (b), and twisting between graphene layers (c).**

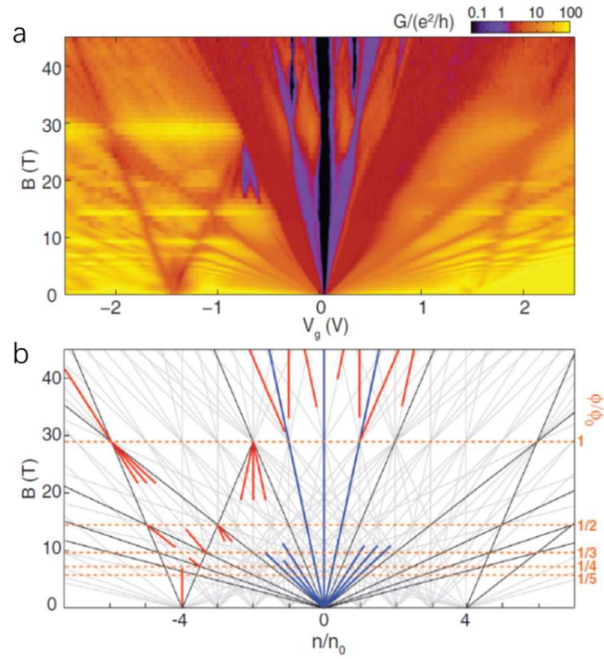

**Fig. 3 Fractal Hofstadter butterfly revealed in a color mapping of conductance to magnetic field B and gate voltage Vg (a) and corresponding Wannier diagram(b) in graphene/hBN superlattice, reproduced from (Hunt, *et al.*, 2013).**

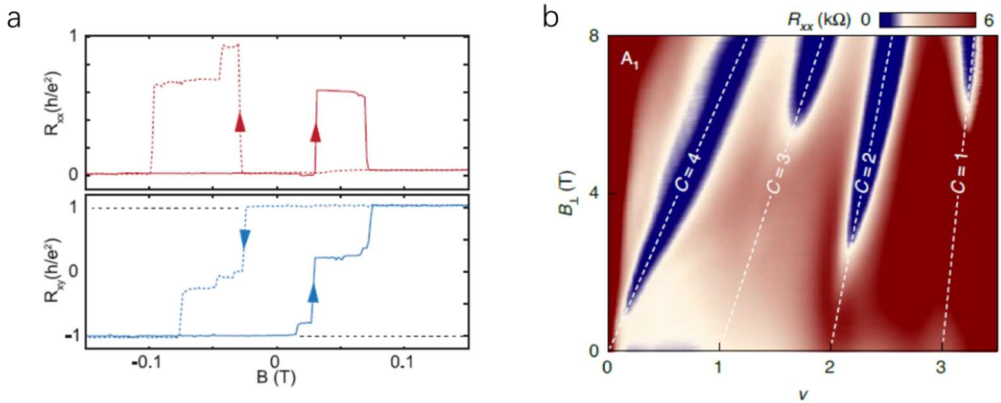

**Fig. 4 (a) Observation of quantized anomalous Hall effect in hBN aligned TBG, i.e. $R_{xy}=h/(Ce^2)$ and $R_{xx}=0$ $h/e^2$ with Chern number C=1, reproduced from (Serlin, *et al.*, 2020). (a) Observation of sequential filled Chern insulator with C=4-*v*, reproduced from (Das, *et al.*, 2021)**